\documentclass[showpacs, oneside, twocolumn, prd, amsmath, amssymb, nofootinbib, %superscriptaddress
]{revtex4}
\usepackage{cases}
\usepackage{amsmath}
\usepackage{amssymb}
\usepackage{amsfonts}
\usepackage{amssymb}
\usepackage{dcolumn}
\usepackage{bm}
\usepackage{bbm}
\usepackage{graphicx}
\usepackage{xcolor}
\usepackage{array}
\usepackage{subfigure}
\usepackage{hyperref}

\newcommand{\be}{\begin{equation}}
\newcommand{\ee}{\end{equation}}
\newcommand{\ba}{\begin{eqnarray}}
\newcommand{\ea}{\end{eqnarray}}
\newcommand{\no}{\nonumber \\}
\newcommand{\gsim}{\mathrel{\hbox{\rlap{\lower.55ex \hbox {$\sim$}}
                   \kern-.3em \raise.4ex \hbox{$>$}}}}
\newcommand{\lsim}{\mathrel{\hbox{\rlap{\lower.55ex \hbox {$\sim$}}
                   \kern-.3em \raise.4ex \hbox{$<$}}}}

\def\roughly#1{\mathrel{\raise.3ex\hbox{$#1$\kern-.75em%
\lower1ex\hbox{$\sim$}}}}
\def\lsim{\roughly<}
\def\gsim{\roughly>}

\def\({\left(}
\def\){\right)}
\def\[{\left[}
\def\]{\right]}
\def\<{\langle}
\def\>{\rangle}
\def\pd{\partial}

\def\l{{\lambda}}

\def\d{{\delta}}

\def\o{{\omega}}

\def\g{{\gamma}}

\def\h{{\eta}}

\def\m{{\mu}}

\def\r{{\rho}}
\def\s{{\sigma}}

\def\t{{\tau}}

\def\ps{{\psi}}
\def\Ph{{\Phi}}
\def\Ps{{\Psi}}

\newcommand{\omits}[1]{}

\newcommand{\cN}{{\cal N}}
\newcommand{\cA}{{\cal A}}

\hypersetup{colorlinks=true,
            breaklinks=true,
            pdfstartview=Fit,
            linkcolor=blue,
            citecolor=blue,
            urlcolor=blue}

\usepackage{color}

\begin{document}

\title{Holographic Preheating: Quasi-Normal Modes and Holographic Renormalization}

\author{
Yi-Fu Cai$^{2}$\footnote{yifucai@ustc.edu.cn},
Shu Lin$^{1}$\footnote{linshu8@mail.sysu.edu.cn},
Junyu Liu$^{2,3}$\footnote{jliu2@caltech.edu},
Jia-Rui Sun$^{1}$\footnote{sunjiarui@sysu.edu.cn}}

%\author{Yi-Fu Cai$^{2}$} \footnote{yifucai@ustc.edu.cn}
%\author{Shu Lin$^{1}$} \footnote{linshu8@mail.sysu.edu.cn}
%\author{Junyu Liu$^{2, 3}$} \footnote{jliu2@caltech.edu}
%\author{Jia-Rui Sun$^{1}$} \footnote{sunjiarui@sysu.edu.cn}

\affiliation{1) School of Physics and Astronomy, Sun Yat-Sen University, Guangzhou 510275, China}
\affiliation{2) CAS Key Laboratory for Researches in Galaxies and Cosmology, Department of Astronomy, University of Science and Technology of China, Hefei, Anhui 230026, China}
\affiliation{3) Walter Burke Institute for Theoretical Physics, California Institute of Technology, Pasadena, CA 91125, United States}

\begin{abstract}
In the holographic description of cosmic preheating proposed in an accompanied Letter \cite{Cai:2016}, the energy transfer between the inflaton and matter field at strong coupling is suggested to be mimicked by superfluid and normal components of a superconductor on Friedmann-Robertson-Walker (FRW) boundary in an asymptotically Anti-de Sitter (AdS) spacetime. In this paper we investigated two aspects of the scenario of holographic preheating that are not included in the accompanied work. Firstly, we study in detail the evolution of the quasi-normal modes (QNMs) surrounding a metastable hairy black hole. This analysis can quantitatively describe the preheating process of the matter field that is produced continuously in the case of strong coupling. Secondly, we present a detailed analysis of the holographic renormalization for the AdS-FRW background, which allows us to extract operator expectation values for studying cosmological implications.
%, which can be transformed from the Schwarzschild-AdS. %This analysis guarantees that the cosmological interpretations of holographic preheating would not be spoiled by high order quantum corrections.
\end{abstract}

\pacs{98.80.Cq, 11.25.Tq, 04.50.Gh, 04.30.Nk}

\maketitle

\section{Introduction}

Our universe is believed to have experienced a period of inflationary phase in the very early times during which its size was enlarged by a factor of about $10^{80}$. During inflation all unwanted primordial relics produced from the Big Bang can be washed out \cite{Guth:1980zm, Brout:1977ix, Starobinsky:1980te, Sato:1980yn, Fang:1980wi}. Within this cosmological paradigm, primordial density fluctuations of quantum origin can be causally generated and then can provide the seeds for the formation of the large-scale structure (LSS) and the Cosmic Microwave Background (CMB) anisotropies \cite{Mukhanov:1990me}. After inflation, it is important to dynamically release the energy stored in the potential of the inflaton field to produce other matter fields so that the universe can reach thermal equilibrium. This process is dubbed as the reheating phase.

The energy transfer from the inflaton field to regular matter was initially studied using the perturbation theory at first order \cite{Abbott:1982hn, Dolgov:1982th, Albrecht:1982mp}. Such an analysis does not take into account the fact that the inflaton field at the end of inflation is in a highly homogeneous condensate state rather than an assembly of free perturbative quanta. As was pointed out in \cite{Dolgov:1989us, Traschen:1990sw}, during the energy transfer from the inflaton to matter field, which are usually modelled by two scalar fields, there exists a sizeable parameter space that contains the parametric resonance instability even if their coupling is weak. This instability was soon applied to realize a phase of preheating in detail in \cite{Kofman:1994rk, Shtanov:1994ce, Kofman:1997yn}, during which the energy transfer from the inflaton to matter fields becomes very dramatic. Similar results were found in alternative paradigms to inflation, such as in bounce cosmology \cite{Cai:2011ci}. For recent reviews of cosmic preheating, we refer to \cite{Bassett:2005xm, Allahverdi:2010xz, Amin:2014eta}.

The key aspect of the present paradigm of cosmic preheating at weak coupling is that the inflaton field oscillations after inflation can make a periodical contribution to the effective mass term of the matter field, which is often described the well-known Floquet theory \cite{McLachlan:1947}. As a result, this weakly coupled cosmic preheating, when combined with observations, can be applied to constrain cosmological parameters of the very early universe even if the coupling between the inflaton and matter field is weak. Namely, the post-inflationary background parameters were studied with a perturbative treatment in Refs. \cite{Podolsky:2005bw, Martin:2010kz, Adshead:2010mc, Mielczarek:2010ag, Easther:2011yq, Dai:2014jja, Martin:2014nya, Cook:2015vqa, Munoz:2014eqa, Cai:2015soa}; and the evolutions of primordial fluctuations during preheating were investigated in Refs. \cite{Finelli:2000ya, Bassett:1999cg, Brandenberger:2007ca, Brandenberger:2008if, Moghaddam:2014ksa, Gong:2015qha, McDonough:2016xvu}. These works have shown that the study of cosmic preheating can play an important role in placing constraints on the cosmology of the very early universe, such as inflationary models.

\begin{figure}
\begin{center}
\includegraphics[width=0.4\textwidth]{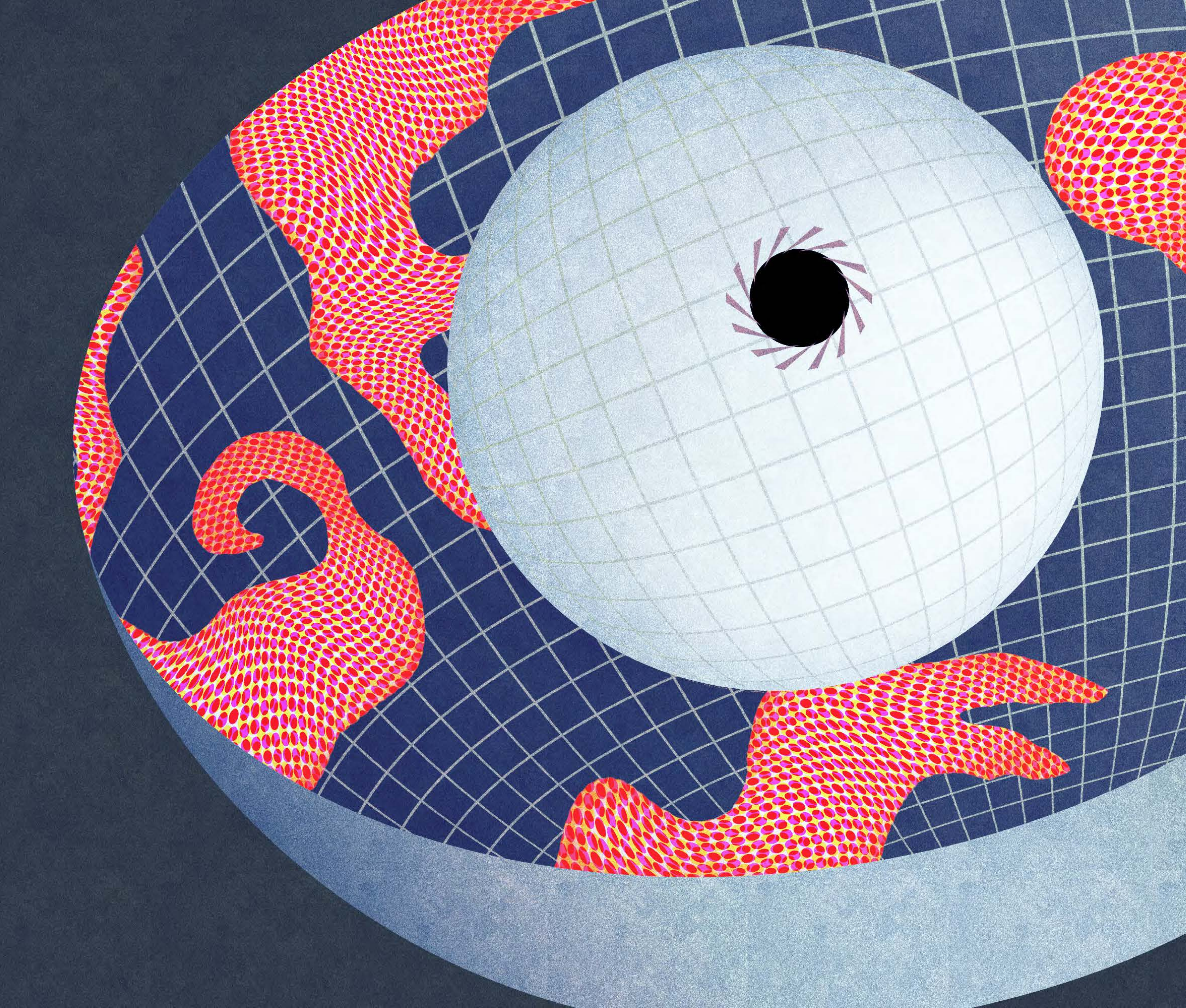}
\end{center}
\caption{Artistic illustration of the {\it holographic preheating} paradigm. In this paradigm, the cosmic preheating in a 4D FRW universe is holographically described by the phase transition in a 5D AdS-Schwarzschild black hole, which is conformally mapped to a AdS-FRW background. From \cite{Cai:2016}.
(credit to Mr. Yulin Hu)
}\label{illustration}
\end{figure}

However, very few study was to address the cosmic preheating at strong couplings, which was expected to occur in some early universe models at extremely high energy scales, namely, see \cite{GarciaBellido:1997mq}.
Study of cosmic preheating at strong coupling enables us to gain more insights on more generic paradigm about this process. Recently, a new description of cosmic preheating at strong coupling was proposed by virtue of the holographic superconductor model in \cite{Cai:2016}. In this mechanism, the preheating process of the inflaton strongly coupled with a class of fermionic matter field can be mimicked by superfluid and normal components of a superconductor, which is holographically described by a five dimensional metastable hairy black hole in the Anti-de Sitter (AdS) spacetime. In order to accommodate with an expanding universe, we take the boundary of this background to be of the Friedmann-Robertson-Walker (FRW) form. This scenario is manifestly different from many other holographic models of cosmology in which the universe is the bulk spacetime, and accordingly, there is a dual quantum field theory on the boundary, as were widely studied in the literature \cite{Chu:2006pa, Das:2006dz, Das:2006pw, Awad:2009bh, Hertog:2005hu, Craps:2007ch, McFadden:2009fg, Brandenberger:2016egn, Ferreira:2016gfg}. The bulk spacetime adopted in \cite{Cai:2016} is then a AdS-FRW and can be mapped to the well-known AdS-Schwarzschild black hole \cite{Apostolopoulos:2008ru} (see also \cite{Lamprou:2011sa, Tetradis:2009bk, Nojiri:2002hz, Camilo:2016kxq}) via coordinate transformation. A similar picture addressing the reheating through holographic thermalization was studied in \cite{Kawai:2015lja}. We refer to Fig. \ref{illustration} for artistic illustration of holographic preheating in a 4D FRW universe.

One goal of the present paper is to study in detail the evolution of the quasi-normal modes (QNMs) of a metastable hairy black hole.
With this analysis, one is able to characterize the process of cosmic preheating at strong coupling quantitatively. Among all QNMs, there always exists one mode of which the imaginary part of the frequency becomes positive, and hence, can drive the cosmological background evolving from the metastable state, which duals to the inflaton-dominant state on the boundary, to the stable state, which duals to the boundary state dominated by the matter field. This can be read by performing a coordinate transformation from the AdS-Schwarzschild geometry to the AdS-FRW one. Thus, the second goal of this paper is to provide a detailed analysis of the holographic renormalization within the AdS-FRW background.
One can see that proper renormalization of the theory requires addition of a curvature term to the FRW boundary.
By taking the derivatives of the renormalized action with respect to the sources, one can obtain relevant vacuum expectation values of the operators, one of which represents for the evolution of the density of the matter field produced during preheating.

The rest parts of the paper are organized as follows. In Section II we introduce the holographic superconductor model with a flat boundary. The background geometry of this model is a meta-stable hairy black hole in the probe limit. Then in Section III, we proceed with the calculation of QNM analysis under this background, which are conformally related to the unstable modes on the FRW boundary. Afterwards, we do the conformal transformation in Section IV, and then, perform the holographic renormalization within the AdS-FRW background. Section V is devoted to a conclusion of our analysis with a discussion.

\section{Holographic model of cosmic preheating}

Our goal is to describe the dynamical evolution of the cosmic system in terms of an inflaton field and a second matter field at strong coupling. The initial state is dominated by the inflaton. The evolution diminishes the inflaton field and amplifies the matter field. The end moment is expected to be a matter dominated phase.

Following the proposal of Ref. \cite{Cai:2016}, we describe the scenario of cosmic preheating by adopting a holographic superconductor type model as follows \cite{Franco:2009yz, Gubser:2008px, Hartnoll:2008kx, Hartnoll:2008vx}:
\begin{align} \label{action}
 S = & \int d^5x\sqrt{-g} \Big[ -\frac{1}{4}F_{MN}F^{MN} -\frac{1}{2}\pd_M\Ps\pd^M\Ps -\frac{1}{2}m^2\Ps^2 \no
 & -\Ps^n\(\pd_Mp-A_M\)\(\pd^Mp-A^M\) \Big],
\end{align}
where we have introduced $F_{MN} \equiv \pd_M A_N-\pd_N A_M$, which is the field strength of the $U(1)$ gauge field in the bulk spacetime. In this action, a complex scalar $\Ps$ is introduced as dual to the inflaton field, and, the $U(1)$ gauge field $A_M$ is introduced as dual to the fermionic matter field. Accordingly, in this scenario, the inflaton field is assumed to be converted to fermionic fields only. We note that the building up of fermionic matter fields would eventually affect the expansion of the universe through the backreaction to the metric. We would like to examine this issue by including the generation of cosmological fluctuations in a forthcoming work \cite{Cai:2017}. Moreover, the model parameter $m$ represents the mass of $\Ps$ and $n$ is a parameter in the coupling between the inflaton and the matter field, which is found to be related with the order of phase transition \cite{Cai:2016} (see also \cite{Franco:2009yz} for the analyses of phase transitions in case of holographic superconductors).

The model is required to be embedded in the background geometry with an FRW boundary, so that it can be applied to describe the expanding universe such as ours. In the probe limit, such a background with an FRW boundary can be obtained by performing a coordinate transformation on an AdS-Schwarzschild black hole \cite{Apostolopoulos:2008ru}. With this process, one can significantly simplify the calculation, since we can first study the dynamics of $\Ph$ and $A_M$ in the AdS-Schwarzschild background, and then, we can translate the results to the background of a thermally expanding universe by performing a coordinate transformation.

The aforementioned model is embedded in the $(d+1)$-dimensional planar AdS-Schwarzschild black hole background,
\begin{align}\label{AdS_Schw}
 ds^2 = -f(r)dt^2 +\frac{dr^2}{f(r)}+\frac{r^2}{L^2}dx_i^2 ~,
\end{align}
where the metric factor is given by $f(r)=\frac{r^2}{L^2}-\frac{M^{d-2}}{r^{d-2}}$, and, $i=1,2,\ldots, d-1$. The vanishing condition of the metric factor yields the radius of the horizon to be
\begin{eqnarray}
 r_H=M^{\frac{(d-2)}{d}}L^{\frac{2}{d}} ~,
\end{eqnarray}
and in the probe limit, the corresponding temperature is related with the horizon via
\begin{align}
 T = \frac{r_Hd}{4\pi L^2} ~.
\end{align}
We set the AdS radius $L=1$ from now for simplicity.

Note that the thermodynamics of the system is only a function of the chemical potential (or equivalently density) of matter field. In order to study thermodynamics, we turn on $\Ps$ and $A_t\equiv\Ph$, of which the dynamics can be derived by solving the background equations of motion. It is convenient to use the dimensionless radial coordinate
\begin{align} \label{zpara}
 z=\frac{r_H}{r} ~,
\end{align}
to express the equations of motion, of which the forms are given by
\begin{align}\label{eom_bg}
 &\frac{\pd^2\Ps}{\pd z^2} +\frac{(1-d-z^d)}{z(1-z^d)}\frac{\pd\Ps}{\pd z} = \frac{m^2 \Ps}{z^2 (1-z^d)} - \frac{n \Ph^2 \Ps^{n-1}}{r_H^2 (1-z^d)^2} ~, \no
 &\frac{\pd^2\Ph}{\pd z^2} +\frac{(3-d)}{z}\frac{\pd\Ph}{\pd z} = \frac{2 \Ps^n \Ph}{z^2 (1-z^d)} ~.
\end{align}
If one solve the above equations near the AdS boundary with $z=\epsilon\to 0$ (where $\epsilon$ is the cutoff which is related to the ultra-violate cutoff of the dual field theory), $\Ps$ and $\Ph$ can be expressed in the following asymptotic forms \footnote{We note that, for $d=4$, the expansions contain additional logarithmic terms.}
\begin{align}
 &\Ps = \Ps_+\epsilon^{\Delta_+}r_H^{-\Delta_+} +\Ps_-\epsilon^{\Delta_-}r_H^{-\Delta_-} +\cdots, \no
 &\Ph = \m-\frac{\r}{r_H^{d-2}}\epsilon^{d-2} +\cdots ~,
\end{align}
where
\begin{align}\label{Delta_pm}
 \Delta_\pm = \frac{d\pm\sqrt{d^2+4m^2}}{2} ~.
\end{align}

As was reported in \cite{Cai:2016}, for the purpose of cosmic preheating in a realistic universe, the dimension parameter $d$ is taken to be $d=4$. Also one can choose $m^2=-3$ (which satisfies the Breitenlohner-Freedman bound in the AdS$_5$ spacetime) such that $\Delta_+=3$ and $\Delta_-=1$. In this case, both $\Ps_+$ and $\Ps_-$ are normalizable modes, which correspond to the {\it vev} and the source of the inflaton field, respectively. Moreover, the value of $\Ph$ on the boundary yields the chemical potential $\mu$ for the fermionic matter field. Thus, one can learn from the study of holographic superconductors \cite{Franco:2009yz} that, a phase transition is present at a critical value for the chemical potential at $\mu_c$ for a given temperature. The phase above $\mu_c$ is an AdS hairy black hole with $\Ps$ condensation, which is dual to inflaton dominated phase. The phase below $\mu_c$ is AdS-Schwarzschild black hole with no $\Ps$ condensation, thus is dual to the matter dominated phase with a vanishing {\it vev} of the inflaton.
The order of this phase transition depends on the value of $n$ which appeared in the last term of the action \eqref{action}, and in our case with $n=3$, the phase transition is of first order. This implies that the inflaton dominated phase extends to the region below $\mu_c$, where the phase becomes metastable.

In the following we will use such a metastable phase as the initial state of cosmic preheating, which is thermodynamically unstable. As we will see from the analysis, such an instability can drive the evolution of the background system under consideration from the inflaton-dominant phase towards a phase dominated by the matter field.

\section{Quasi-Normal Modes of the Metastable State}

In this section, we will focus on the properties of the QNMs generated during the initial stage of the background evolution. The full analysis of the late time evolution should also include the back-reactions, and we would like to address this part in our forthcoming work \cite{Cai:2017}.

Now we are ready to study the QNMs of the metastable phase. The procedure is standard \cite{Amado:2009ts, Kovtun:2005ev} (see also examples for QNMs in metastable phase \cite{Gursoy:2013zxa, Janik:2016btb}). First of all, we turn on the following perturbations:
\begin{align} \label{pert}
 & \Ps \to \Ps(r) +\s(t,\,x) ~,~ p \to \frac{\h(r,t,x)}{\Ps(r)} ~, \no
 & A_t \to \Ph(r) +a_t(r,t,x) ~,~ A_x \to a_x(r,t,x) ~.
\end{align}
The first three modes, which are: $\s$, $\h$ and $a_t$, can compose of a minimum set for analyzing the dynamical system of the inflaton and matter field.

Here we have chosen the axial gauge: $A_r=0$. It is also possible to eliminate $p$ in favor of $A_r$. Without loss of generality, we include the dependence on $x_1\equiv x$ in order to study the effect of spatial inhomogeneity, which requires that one to include $A_x$ in the perturbation modes. It is straightforward to write down the background equations of motion as follows,
\begin{align} \label{pert_eom}
 & \pd_M (\sqrt{-g}\pd^M\Ps ) / \sqrt{-g} = m^2\Ps + 3 \Ps^2 ( \pd_N p-A_N ) ( \pd^N p-A^N ) ~, \no
 & \pd [ \sqrt{-g}\Ps^3 (\pd^N p -A^N ) ] = 0 ~, \no
 & \pd_M ( \sqrt{-g}F^{MN} ) / \sqrt{-g} = 2 \Ps^3 (\pd^N p-A^N ) ~,
\end{align}
in order to derive the perturbation equations.

Afterwards, we insert the perturbation modes introduced in Eq. \eqref{pert} into the above equations, and then, perform the Fourier transformation by taking the form $e^{-i\o t+ikx}$. As a result, we can obtain the perturbation equations list in the following,
%\begin{widetext}
\begin{align}\label{eom_pert}
 & {\omega}^{2} \eta {\Psi}^{2} -{k}^{2} r^{-2} f\eta {\Psi}^{2} - 3i \omega \sigma \Phi {\Psi }^{2} - i \omega {\Psi}^{3} {a}_{t} \no
 & - i k r^{-2} f{\Psi}^{3}{a}_{x} + 3r^{-1} {f}^{2} {\Psi}^{2} \pd_r\eta + f{\Psi}^{2} \pd_r f \pd_r\eta \no
 & -3 {r}^{-1} {f}^{2} \eta \Psi \pd_r \Psi - f\eta \Psi \pd_r f\pd_r \Psi + {f}^{2} \Psi \pd_r\eta \pd_r\Psi \no
 & - f^2\eta\(\pd_r\Psi\)^{2} + {f}^{2}{{\Psi }^{2}}\pd_r^2\eta - {{f}^{2}}\eta\Psi \pd_r^2\Psi=0 ~, \\
 & {\omega}^{2} \sigma -{{k}^{2}}r^{-2} f\sigma -{{m}^{2}} f\sigma +6i \omega\eta\Phi\Psi +6 \sigma {{\Phi }^{2}}\Psi \no
 & +6 \Phi {{\Psi }^{2}}{{a}_{t}} +3 r^{-1} f^2\pd_r\sigma + f\pd_rf \pd_r\sigma + {{f}^{2}}\pd_r^2\sigma =0 ~, \\
 & 2i \omega\eta{{\Psi}^{2}} +6 \sigma \Phi \Psi^2 -k^2 r^{-2} a_t +2\Psi^3 a_t -k r^{-2} \omega a_x \no
 & +3r^{-1} f \pd_r a_t +f \pd_r^2 a_t =0 ~, \\
 & 2ikrf\eta \Psi^2 -kr\omega a_t -r \omega^2 a_x -2rf \Psi^3 a_x - f^2 \pd_r^2 a_x \no
 & - rf \pd_r f\pd_r a_x -r f^2 \pd_r^2 a_x =0 ~, \\
 & 2 f{{\Psi}^{2}} \pd_r \eta -2 f \eta \Psi \pd_r \Psi +i \omega\pd_r{{a}_{t}} +ikr^{-2}f\pd_r{{a}_{x}} =0 ~.
\end{align}
%\end{widetext}
%\textbf{where the prime denotes the derivative with respect to ?}.

It is convenient to rewrite the perturbation equations in terms of the dimensionless coordinate $z$ defined in Eq. \eqref{zpara}.
Then, to solve the dynamics of QNMs, we impose the ingoing wave condition on the horizon $z=1$.
This leads to the asymptotic solutions near the horizon as follows,
\begin{align}\label{horizon_sol}
 &\h = (1-z)^{-\frac{i\o}{4r_H}} [ \h_0 +\h_1(1-z) +\cdots ] ~, \no
 &\s = (1-z)^{-\frac{i\o}{4r_H}} [ \s_0 +\s_1(1-z) +\cdots ] ~, \no
 &a_t = (1-z)^{1-\frac{i\o}{4r_H}} [ a_{t0} +a_{t1}(1-z) +\cdots ] ~, \no
 &a_x = (1-z)^{-\frac{i\o}{4r_H}} [ a_{x0} +a_{x1}(1-z) +\cdots ] ~.
\end{align}
Note that, there are only three free parameters in \eqref{horizon_sol}. We can parameterize the solutions by $\h_0$, $\s_0$ and $a_{x0}$, with the remaining parameters determined from Eq. \eqref{eom_pert}, and these three parameters can lead to three independent solutions. In the following, we label the corresponding solutions by the subscripts $I$, $II$ and $III$, which are given by:
\begin{align}
 &\h_0^I=1 ~,~ \s_0^I=0 ~,~ a_{x0}^I=0 ~, \no
 &\h_0^{II}=0 ~,~ \s_0^{II}=1 ~,~ a_{x0}^{II}=0 ~, \no
 &\h_0^{III}=0 ~,~ \s_0^{III}=0 ~,~ a_{x0}^{III}=1 ~,
\end{align}
respectively.
Moreover, there remains a pure gauge solution, which can be generated from the gauge transformation on a trivial solution as follows,
\begin{align}\label{pure}
\h^{IV}=i\l\Ps ~,~ \s^{IV}=0 ~,~ a_t^{IV}=\o\l ~,~ a_x^{IV}=-k\l ~,
\end{align}
and has been labeled by the subscript $IV$.

For a generic solution constructed from linear combination of the above four basis solutions, the asymptotic expansions of the QNMs near the boundary then take the forms to be
\begin{align}\label{generic_asym}
 &\h=\h_bz+\cdots ~,~ \s=\s_bz^3+\cdots ~, \no
 &a_t=a_{tb}+\cdots ~,~ a_x=a_{xb}+\cdots ~,
\end{align}
The leading coefficients are dual to sources to the corresponding fields on the Minkowski boundary. The conditions of QNMs require that the modes live without external sources, and thus, all leading coefficients should vanish. This can be formulated as the condition on the above basis solutions as follows,
\begin{align}\label{det_qnm}
\begin{vmatrix}
\h_b^I& \h_b^{II}& \h_b^{III}& \h_b^{IV}\\
\s_b^I& \s_b^{II}& \s_b^{III}& \s_b^{IV}\\
a_{tb}^I& a_{tb}^{II}& a_{tb}^{III}& a_{tb}^{IV}\\
a_{xb}^I& a_{xb}^{II}& a_{xb}^{III}& a_{xb}^{IV}
\end{vmatrix}=0.
\end{align}

Note that Eq. \eqref{det_qnm} is independent of the normalization of solutions, and hence, we can set $\l=1$ in the pure gauge solution \eqref{pure}. The condition allows us to determine the QNMs for a fixed wave number $k$. The QNMs with the positive imaginary part correspond to the existence of an instability for the metastable state, which can be clearly read by inserting this imaginary part into the exponent factor $e^{-i\o t+ikx}$. The real part of QNMs indicate that there also exist oscillations during the amplification of matter field.

\begin{figure}
\includegraphics[width=0.38\textwidth]{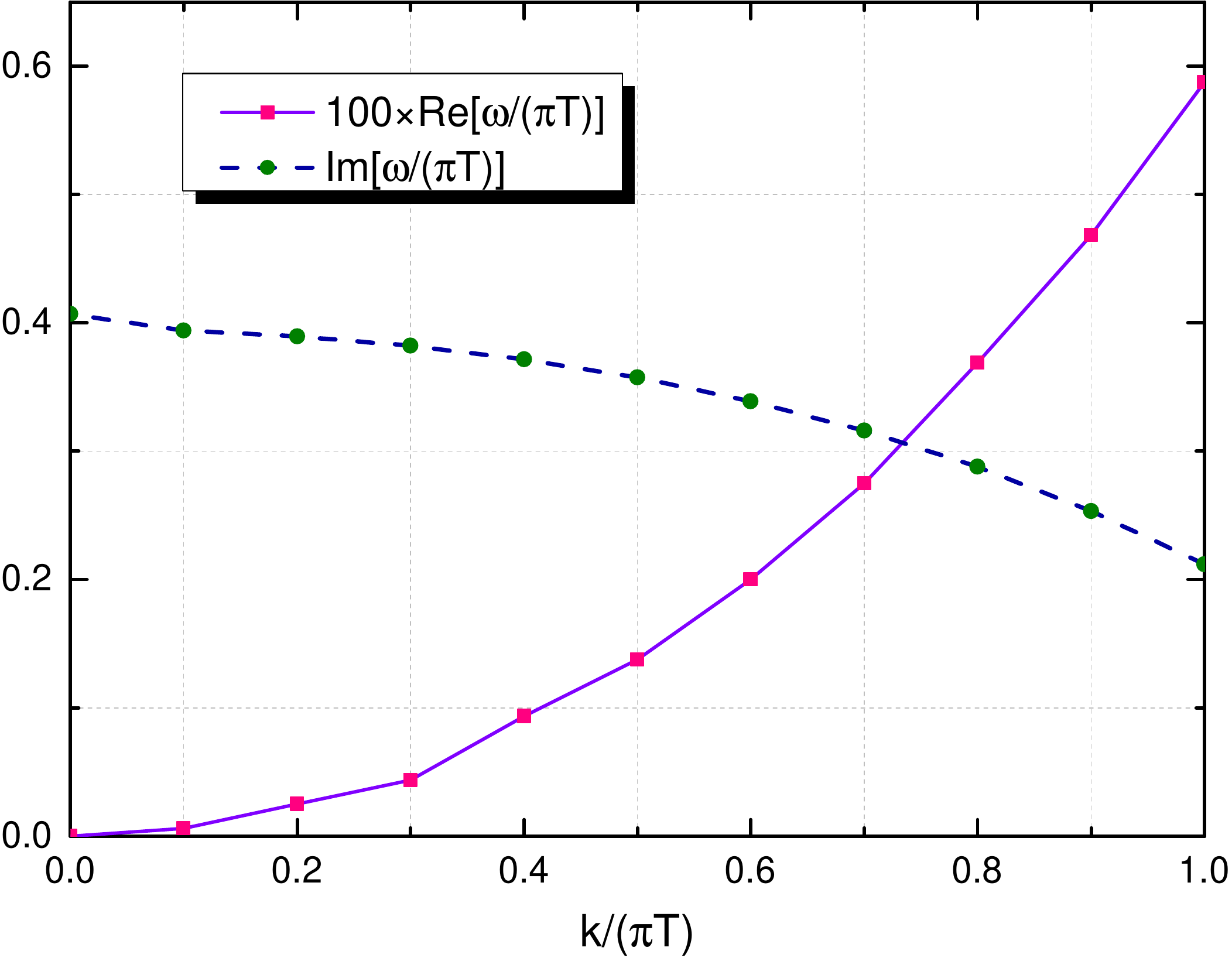}
\includegraphics[width=0.40\textwidth]{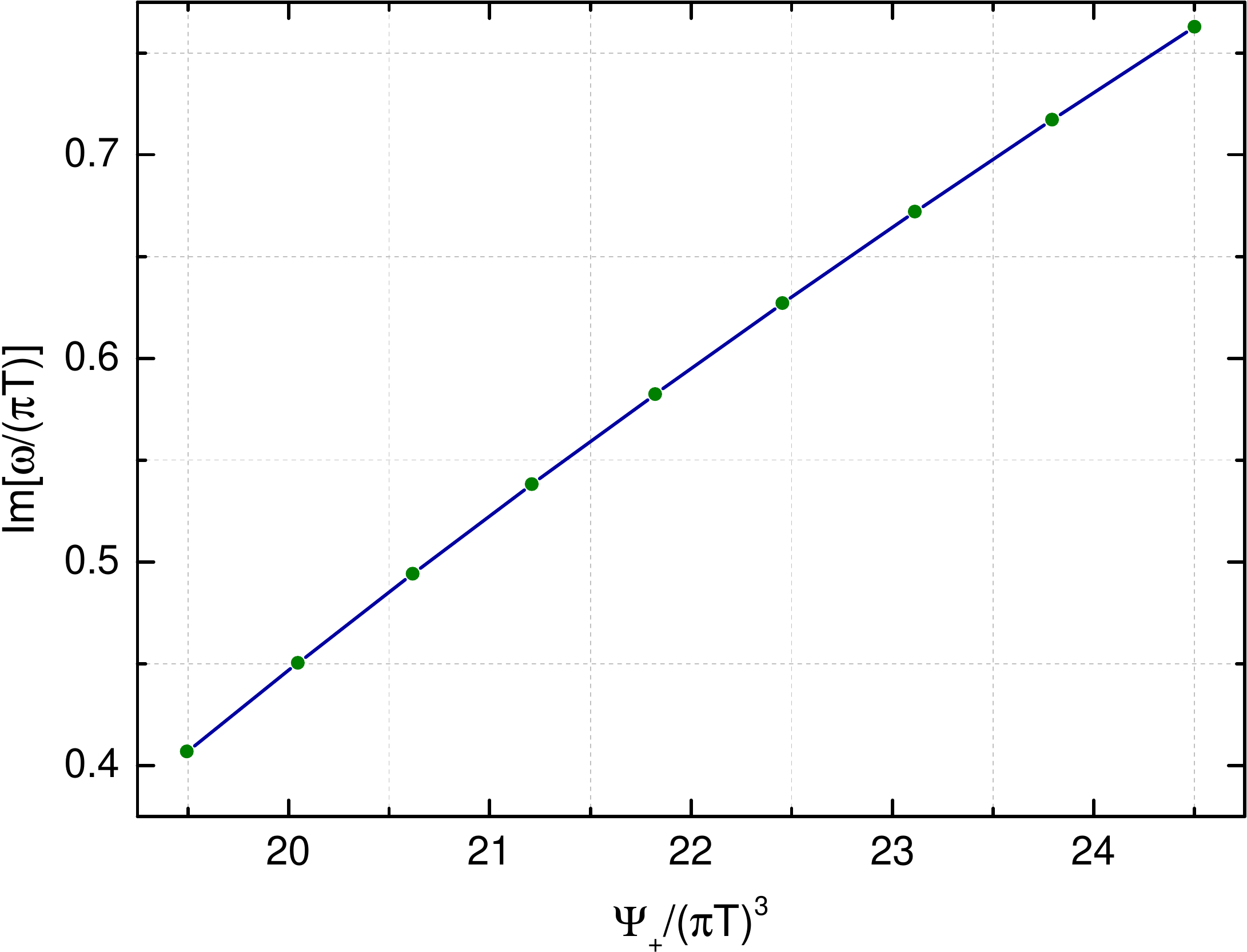}
\caption{Upper: The real and imaginary parts of the frequency of QNMs at $T=0.2$ as functions of the wave number $k$ for a fixed inflaton's {\it vev} $\frac{\Ps_+}{(\pi T)^3}=19.5$. We see that the most unstable QNM (one with the largest imaginary part) occurs at $k=0$, where the real part vanishes. Lower: The imaginary part of the most unstable mode as a function of dimensionless {\it vev} of the inflaton $\frac{\Psi_+}{(\pi T)^3}$ with $k=0$. Note that the phase transition is of first order, so the order parameter $\Ps_+$ does not extend to zero at the critical point.}\label{qnm}
\end{figure}

The dependencies of QNMs on the wave number $k$ and the {\it vev} of the inflaton are numerically shown in Fig. \ref{qnm}. From the upper panel of Fig. \ref{qnm}, one can see that at the small $k$ regime, the amplification of matter field is the most efficient, which corresponds to the most dramatic regime during the preheating process. This property is very similar to the case of cosmic preheating at weak coupling in which the exponential growth can be realized when the coupling satisfies the requirement of parametric resonance, except that the growth in the strong coupling case is steady continuous without additional requirement on the couplings. Moreover, at the large $k$ regime, the oscillations become more and more important to describe the evolutions of these QNMs. According to the curve in the lower panel of the figure, we also learn that a larger value of the inflaton's {\it vev} tends to trigger a faster conversion to the matter field, with the imaginary part of the frequency being an approximately linear function in the {\it vev} of the inflaton.

\section{Holographic renormalization in a AdS-FRW background}

In the previous section we have investigated the QNMs of the metastable state in the AdS-Schwarzschild black hole background. After that, we need to map our results to the FRW-AdS background. Note that, The FRW-AdS background can be obtained from the AdS-Schwarzschild black hole via a coordinate transformation \cite{Apostolopoulos:2008ru}, and its application to holographic preheating has been performed in \cite{Cai:2016}. In this section, we present the holographic renormalization in the AdS-FRW background to ensure that the cosmological implications derived in Ref. \cite{Cai:2016} are reliable.

By adopting the Fefferman-Graham coordinates, the 5-dimensional AdS-FRW spacetime can be expressed as
\begin{eqnarray}\label{AdS_FRW}
 ds^2 = \frac{1}{\xi^2} \Big[ d\xi^2 -\cN(\t,\xi)^2d\t^2 +\cA(\t,\xi)^2 d\vec{x}^2 \Big] ~.
\end{eqnarray}
with $d\vec{x}^2 \equiv dx_1^2 + dx_2^2 + dx_3^2$.
Its boundary is located at $\xi \to 0$, which corresponds to a 4-dimensional FRW universe with $\cN \to 1$ and $\cA \to a$. Here, $a$ is the scale factor of the universe living on the boundary. The expressions of $\cN$ and $\cA$ are give by,
\begin{align}\label{cAN}
 & \cA^2 = a^2 -\frac{{\dot a}^2}{2}\xi^2 +\frac{{\dot a}^4 +4r_H^4}{16a^2}\xi^4 ~, \no
 & \cN = \frac{{\dot \cA}}{{\dot a}} ~,
\end{align}
where the dot represents derivative with respect to $\t$.

The AdS-Schwarzschild metric and the AdS-FRW one are related through a set of coordinate transformations, which maps $(t,\,r)$ to $(\t,\,\xi)$ through the following relations,
\begin{align}\label{coord_trans}
 r = \frac{\cA}{\xi} ~,~ {\dot t} = -\frac{{\dot \cA}r'}{f{\dot a}} ~,~ t'=-\frac{{\dot a}}{\xi f} ~,
\end{align}
and here, the prime denotes the derivative with respect to $\xi$.
The explicit form of $t(\t,\,\xi)$ can be obtained by perform an integral on \eqref{coord_trans}. The coordinate transformations reduce to the conformal transformation on the AdS boundary, and it relates the stress tensor in the FRW background to that in the Minkowski \cite{Apostolopoulos:2008ru}.

We have chosen $m^2=-3$ such that the indices of $\Ps$ take values in \eqref{Delta_pm}. Then, we work in the axial gauge $A_\xi=0$. It follows from the equations of motion \eqref{pert_eom} that the asymptotic expansions of the fields are given by
\begin{align}\label{asym_zb}
 &A_\t = a_{\t0} +a_{\t1}\xi +a_{\t2}\xi^2 +a_{\t h} \xi^2\ln\xi +\cdots ~, \no
 &A_x = a_{x0} +a_{x1}\xi +a_{x2}\xi^2 +a_{xh} \xi^2\ln\xi +\cdots ~, \no
 &\Ps = \Ps_1\xi +\Ps_2\xi^2 +\Ps_3\xi^3 +\Ps_h\xi^3\ln\xi +\cdots ~, \no
 &p = p_0 +p_1\xi +p_2\xi^2 +p_h \xi^2\ln \xi +\cdots ~.
\end{align}
Note that, the coefficients in \eqref{asym_zb} are not all independent. Applying the equations of motion and the following asymptotic expansions for $\cN$ and $\cA$ (see \cite{Apostolopoulos:2008ru}):
\begin{align}\label{NA_exp}
&\cN=1+\frac{{\dot a}^2-2a{\ddot a}}{4a^2}\xi^2-\frac{3r_H^4}{8a^4}\xi^4+\cdots, \no
&\cA=a-\frac{{\dot a}^2}{4a}\xi^2+\frac{r_H^4}{8a^3}\xi^4+\cdots,
\end{align}
we can determine $a_{\t1}=0$, $a_{x1}=0$ and $\Ps_2=0$. Furthermore, $a_{\t h}$, $a_{xh}$, $\Ps_h$ and $p_h$ can be expressed in terms of $a_{\t 0}$, $a_{x0}$, $\Ps_1$ and $a$. We will not show explicit expressions here.
The undetermined coefficients will need further information from the bulk.

Afterwards, we wish to show that these coefficients give rise to the {\it vev} of the corresponding operators. We proceed by calculating the on-shell action as shown below
\begin{align}
 S = & \int_{\pd}-\frac{1}{2}\sqrt{-g}F^{\xi N}A_N+\int\frac{1}{2}\pd_M\(\sqrt{-g}F^{MN}\)A_N \no
 & +\int_{\pd}-\frac{1}{2}\sqrt{-g}\pd^\xi\Ps\Ps+\int\frac{1}{2}\pd_M\(\sqrt{-g}\pd^M\ps\)\Ps \no
 & -\int\frac{m^2}{2}\sqrt{-g}\Ps^2 -\int_{\pd}\sqrt{-g}\Ps^3\pd^\xi p\cdot p \no
 & +\int\pd_M\(\sqrt{-g}\Ps^3\(\pd^Mp-A^M\)\)p \no
 & +\int\Ps^3\sqrt{-g}A_M\(\pd^Mp-A^M\) ~.
\end{align}
Here we has used the short hand notations: $\int=\int d^5x$, $\int_\pd=\int d^4x$, and keep the boundary terms only on the AdS boundary. By using the background equations of motion again, we find the on-shell action as
\begin{align}\label{Sos}
 S_{os} = & \int_\pd -\frac{1}{2}\sqrt{-g} \big( F^{\xi N}A_N+\pd^\xi\Ps\Ps - \Ps^3\pd^\xi p\cdot p \big) \no
 & + \int\frac{3}{2} \Ps^3 \sqrt{-g} \big( \pd^Mp-A^M \big) \big( \pd_Mp-A_M \big) ~.
\end{align}
We note that last term is a bulk term. It is known that this term could lead to the finite scheme-dependent contribution to the on-shell action \cite{Karch:2005ms}. We will drop the bulk term in the following.

The on-shell action contains the divergences in the form of
\begin{align}
 S_{os} = -\frac{a^3\Ps_1^2}{2\xi^2} +\(\cdots\)\ln\xi +\text{finite terms}.
\end{align}
We could add to it the following counter terms
\begin{align}
 & S_{\Ps^2}=\int_\pd\frac{1}{2}\sqrt{-\g}\Ps^2 ~, \no
 & S_{D\Ps^2}=\int_\pd\frac{1}{2}\sqrt{-\g}\(D\Ps\)^2 ~, \no
 & S_{f^2}=\int-\frac{1}{4}\sqrt{-\g}f^2 ~,
\end{align}
where we have used $\g$ to denote the induced metric on a cutoff surface close to the AdS boundary and $D$ is the covariant derivative compatible with $\gamma$. $f$ is the induced field strength on the cutoff surface. We find that the $\frac{1}{\xi^2}$ term can be canceled by the contribution of $S_{\Ps^2}$, and the $\ln\xi$ divergence can be partially canceled by the terms $S_{D\Ps^2}$ and $S_{f^2}$.
We note that the regular forms for $S_{D\Ps^2}$ and $S_{f^2}$ provided below are not enough to cancel all divergences
\begin{align}
 & S_{f^2} = \frac{a}{2} \big[ 2 (\pd_x a_{\t 0}) (\pd_\t a_{x0}) -(\pd_x a_{\t 0})^2 -(\pd_\t a_{x0})^2 \big] +O(\xi) ~, \no
 & S_{D\Ps^2} = \frac{a}{2} \big[ (\pd_x\Ps_1)^2 -a^2(\pd_\t\Ps_1)^2 \big] +O(\xi) ~.
\end{align}
%
%We need to use their alternative forms, which are obtained by integration by parts
%\begin{align}
% & S_{D\Ps^2} = \frac{a}{2} \big[ \Ps_1 ( 3a{\dot a}\pd_\t\Ps_1 +a^2\pd_\t^2\Ps_1 ) -\Ps_1\pd_x^2\Ps_1 \big] +O(\xi) ~, \no
% & S_{f^2} = -\frac{1}{2}\big[a_{\t 0} ( -a\pd_x^2a_{\t 0} +a\pd_\t\pd_x a_{x0} ) -a_{x0}(-{\dot a}\pd_x a_{\t 0}+{\dot a}\pd_\t a_{x0} -a\pd_\t\pd_x a_{\t 0} +a\pd_\t^2 a_{x0}) \no
%&+O(\xi)\big].
%\end{align}
%These terms still can not fully eliminate all the $\ln\xi$ divergence.
%
This is because that there remains one extra contribution not yet taken into account in the above study, which is the boundary term due to the curvature of an FRW metric as shown below,
\begin{align}
 S_{R}=\frac{1}{24}\int_\pd\sqrt{-\gamma}R_\g ~.
\end{align}

In order to further simplify the on-shell action, we also add finite counter terms $\frac{1}{2}S_{D\Ps^2}$ and $\frac{1}{2}S_{f^2}$. The inclusion of these terms corresponds to a specific choice for the scheme. To combine all the above terms together, we finally obtain the renormalized on-shell effective action as follows,
\begin{align}
 S_{ren} =& \int_\pd \Big[ -a a_{x0}a_{x2} +a^3a_{\t 0}a_{\t 2} -a^3\Ps_1\Ps_3 \no
 & +p_0(3a^2{\dot a}a_{\t 2} -a\pd_x a_{x2} +a^3\pd_\t a_{\t 2}) \Big] ~.
\end{align}
Taking derivative with respect to source terms, we eventually obtain the corresponding {\it vev} for all relevant fields, which are given by,
\begin{align}\label{dict_FRW}
 & j^\t = \lim_{\xi\to0} -\frac{\d S_{ren}}{\sqrt{-\gamma}\d a_{\t 0}} = a_{\t 2} ~, \no
 & j^x = \lim_{\xi\to0} -\frac{\d S_{ren}}{\sqrt{-\gamma}\d a_{x0}} = \frac{a_{x2}}{a^2} ~, \no
 & O = \lim_{\xi\to0} -\frac{\d S_{ren}}{\sqrt{-\gamma}\d \Ps_1} = \Ps_3 ~.
\end{align}

Interestingly we note that, the first term in \eqref{dict_FRW} exactly describes the growth of the matter field density in the FRW universe during holographic preheating. Its result has already been reported in the end of \cite{Cai:2016}. The second term in \eqref{dict_FRW} is associated with the inhomogeneity of the produced matter field, which is of interest for cosmological perturbations, but we would like to leave it for future study \cite{Cai:2017}. The third term in \eqref{dict_FRW} describes the evolution of the inflaton field, which in our case exhibits a damping effect.

\section{Conclusion}

Holographic preheating is a newly proposed scenario that addressed the energy transfer from the primordial inflaton field to other matter fields in the situation of strong coupling. This could be in analogy with some particle physics phenomenon, such as, heavy ion collisions \cite{Berges:2008pc}, which may also occur in the early universe when the universe reaches thermal equilibrium \cite{Micha:2004bv, Micha:2002ey}. It has been found in \cite{Cai:2016} that, such a process may be mimicked by a model of holographic superconductor with interacting superfluid and normal components.

In this paper, we have studied two aspects of the holographic preheating paradigm, which are the QNM analysis of a metastable hairy black hole and the holographic renormalization in AdS-FRW background, respectively. For the first part of the study, the QNMs surrounding the hairy AdS-Schwarzschild black hole are related to dynamics on the boundary, which correspond to the exponential matter creations in Minkowski space. By analyzing the QNMs, we can learn the detailed process of cosmic preheating at strong coupling quantitatively. Among all QNMs, there always exists one mode of which the imaginary part of the frequency becomes positive and hence drive the background to evolve from the metastable phase, which is dual to the boundary state dominated by the inflaton, to the stable phase, which is dual to the boundary state dominated by the matter.

In order to grasp the cosmological information correctly, one needs to map the results in Minkowski space to the counterpart in FRW space by conformal transformation. %and then, in the second part of the study, we present the detailed procedure of the holographic renormalization with this background. We found that the addition of a curvature term into the boundary action is required. %This coincides with the phenomenological expectation that an FRW universe could live on the boundary.
%By using the renormalized effective action, we derived the expectation values for relevant fields, which characterize the density of the matter field and the inflaton,
This can be done by a coordinate transformation from the AdS-Schwarzschild geometry to the AdS-FRW one. To translate the results in AdS-FRW to the cosmic system, we need a holographic renormalization procedure. Thus, the second goal of the present paper is to provide a detailed analysis of the holographic renormalization within the AdS-FRW background. One can see from the calculation that, the consistency condition requires addition of a curvature term into the FRW boundary that consists the effective field description of our universe during the holographic preheating. Moreover, by taking the derivatives of the renormalized effective action with respect to the sources, one can obtain the relevant vacuum expectation values of the operators, two of which represent for the growth of the matter field density and the damping of the inflaton field during preheating.

\section*{Acknowledgments}
We are grateful to Robert Brandenberger, Sean Carroll, Hong Lu, Antonino Marciano, Rong-Xin Miao, Wen-Yu Wen and Yi Yang for valuable scientific communications, while Yulin Hu and Rui Niu for valuable artistic communications on the figure.
S.L. thanks the University of Science and Technology of China (USTC) for providing a stimulating environment during the {\it East Asia Joint Workshop on Fields and Strings} and {\it 2016 CfA@USTC Junior Symposium on Gravitation and Cosmology}.
J.L. thanks the hospitality from Sun Yat-Sen University (SYSU) during his visit in the summer of 2016.
Y.F.C. is supported in part by the Chinese National Youth Thousand Talents Program (No. KJ2030220006), by the USTC start-up (No. KY2030000049) and by NSFC (Nos. 11421303, 11653002).
S.L. is supported in part by the Chinese National Youth Thousand Talents Program and by the SYSU Junior Faculty Fund.
J.L. is supported in part by the physics department at Caltech and by the Fund for Fostering Talents in Basic Science of NSFC (No. J1310021).
J.R.S. is supported in part by NSFC (No. 11205058), the Fundamental Research Funds for Central Universities and by the Open Project Program of State Key Laboratory of Theoretical Physics at CAS (No. Y5KF161CJ1).
Part of numerical simulations are operated on the computer cluster LINDA in the particle cosmology group at USTC.

\end{document}